\documentclass[aps,prl,letterpaper,twocolumn,preprintnumbers,floatfix,superscriptaddress]{revtex4}

\usepackage{graphicx}
\usepackage{dcolumn}
\usepackage{bm}
\usepackage{epsfig,cancel}

\usepackage{amsmath}
\usepackage{amssymb}
\usepackage{graphicx,bm}

\def\fun#1#2{\lower3.6pt\vbox{\baselineskip0pt\lineskip.9pt
  \ialign{$\mathsurround=0pt#1\hfil##\hfil$\crcr#2\crcr\sim\crcr}}}

\def\lsim{\mathrel{\rlap{\raise 2.5pt \hbox{$<$}}\lower 2.5pt\hbox{$\sim$}}}
\def\gsim{\mathrel{\rlap{\raise 2.5pt \hbox{$>$}}\lower 2.5pt\hbox{$\sim$}}}

\input epsf

\newcommand{\be}{\begin{equation}}
\newcommand{\ee}{\end{equation}}
\newcommand{\bea}{\begin{eqnarray}}
\newcommand{\eea}{\end{eqnarray}}

\usepackage{color}

\begin{document}

\title{Post-ACME2013 CP-violation in Higgs Physics and Electroweak Baryogenesis}

\author{Ligong Bian}
\affiliation{State Key Laboratory of Theoretical Physics and Kavli Institute for Theoretical Physics China, Institute of Theoretical Physics, Chinese Academy of Sciences, Beijing 100190, P. R. China}

\author{Tao Liu}
\affiliation{Department of Physics, The Hong Kong University of  Science and Technology, Clear Water Bay, Kowloon, Hong Kong}

\author{Jing Shu}
\affiliation{State Key Laboratory of Theoretical Physics and Kavli Institute for Theoretical Physics China, Institute of Theoretical Physics, Chinese Academy of Sciences, Beijing 100190, P. R. China}

\begin{abstract}
We present a class of cancellation mechanisms to suppress the total contributions of Barr-Zee diagrams to the electron electric dipole moment (eEDM). This class of mechanisms are of particular significance after the new eEDM upper limit, which strongly constrains the allowed magnitude of CP-violation in Higgs couplings and hence the feasibility of electroweak baryogenesis (EWBG), were released by the ACME collaboration in 2013. We point out: if both the CP-odd Higgs-photon-photon ($Z$ boson) and the CP-odd Higgs-electron-positron couplings are turned on, a cancellation may occur either between the contributions of a CP-mixing Higgs boson, with the other Higgs bosons being decoupled, or between the contributions of a CP-even and a CP-odd Higgs bosons. With the assistance of the cancellation mechanisms, a large CP-phase in Higgs couplings with viable electroweak baryogenesis (EWBG) is still allowed. The reopened parameter regions would be probed by the future neutron, mercury EDM measurements, and direct measurements of Higgs CP-properties at the LHC and future colliders. 

\end{abstract}

\pacs{draft}

\maketitle

\section{Introduction}

The baryon asymmetry in the Universe (BAU) nowadays, i.e.,~\cite{Ade:2013zuv,Beringer:1900zz}  
\begin{eqnarray}
\frac{n_b}{s}\approx(0.7-0.9)\times10^{-10}  \neq 0
\end{eqnarray}
has puzzled people for more than half a century. Here $s$ is entropy density of the Universe. Among various dynamical mechanisms to solve this puzzle, electroweak baryogenesis (EWBG) falls in the most popular class, due to its potential testability at the Large Hadron Collider and in the other experiments.  A generic feature of the EWBG is that the CP phases employed to generate the cosmic baryon asymmetry need to enter the couplings between the Higgs sector and particles which either exist in the Standard Model (SM)  or are introduced in new physics, no matter the CP-phases are flavor-diagonal, off-diagonal~\cite{Liu:2011jh}, or flavor-decoupled. Otherwise, these CP-phases are decoupled from electroweak phase transition (EWPT) and the EWBG will never be implemented. The measurement of the Higgs CP-properties therefore provides important information to solve the BAU puzzle.

Motivated by this, the CP-properties of the Higgs boson discovered in 2012~\cite{Aad:2012tfa} have been extensively studied by both theorists ~\cite{Cheung:2014oaa,Inoue:2014nva,Shu:2013uua,Bishara:2013vya,Brod:2013cka} and experimental groups~\cite{ATLAS:2012soa} since its discovery~\cite{Aad:2012tfa}, by using a method of direct measurements at the LHC. Given the limited statistics, however, the sensitivity of the LHC at this stage is still low. On the other hand, fast progress has been made in indirect measurements. Using the polar molecule thorium monoxide (ThO), the ACME collaboration reported an upper limit on the eEDM recently~\cite{Baron:2013eja} \footnote{In principle, the CP-odd electron-nucleon interactions can also contribute to the effective electron EDM in a paramagnetic system like Tho. However, those operators are mediated by Higgs bosons and hence highly suppressed by the electron and valence quark masses, thus the ACME results can be safely interpreted as an upper limit for the electron EDM $d_e^E$.   }
\begin{eqnarray}
|d_e|< 8.7\times10^{-29} e{\rm cm} 
\end{eqnarray}
at 90$\%$ confidence level, an order of magnitude stronger than the previous best limit. This limit severely constrains the allowed magnitude of CP-phases in the Higgs couplings
~\cite{Cheung:2014oaa,Inoue:2014nva,Bishara:2013vya,Brod:2013cka} via Barr-Zee diagrams, causing a tension between the observation and the CP-phase required for  successfully implementing EWBG (e.g., see~\cite{Li:2008kz} where the expected projection of the eEDM bounds to the EWBG in the MSSM was studied.)~\footnote{One exception is~\cite{Liu:2011jh}, where the EWBG is driven by flavor off-diagonal CP-phases, with their contributions to the EDM being trivial. In this case, the observables in the $B$- or the other meson systems are potentially more important probes.}. 

In this letter we point out that in these studies a crucial effect was more or less ignored, which can dramatically change the conclusions. This is due to the fact that generally both the CP-odd Higgs-photon-photon and the CP-odd Higgs-electron-positron couplings can be or tend to be turned on. These two couplings contribute to the eEDM separately and simultaneously. If there exists a cancellation between their contributions (as we will show below in two different contexts: the type-II two Higgs Doublet Model (2HDM) where the tree-level CP-phase arises from the pure Higgs sector,  and the Minimal Supersymmetric Standard Model (MSSM) where the tree-level CP-phase arises from Higgs-superparticle interaction sectors.), even if the magnitudes of the CP-phases in Higgs couplings are large, the current ACME bound can be well-satisfied. In such a case, EWBG can still be successfully implemented.

\section{General Analysis}

In an effective Lagrangian for a Higgs sector, the relevant operators are given by:
\bea
{\cal L}_{\rm eff}&=&\frac{m_f}{v}   \sum_i h_i \bar f\left(c_f^i+i\tilde c_f^i \gamma^5\right)f\nonumber \\ &&+\frac{\alpha}{\pi v} \sum_i h_i \left(c_\gamma^i F^{\mu\nu} V_{\mu\nu}
 +\tilde c_\gamma^i F^{\mu\nu}\tilde V_{\mu\nu}\right),
\label{eq:EFT}
\eea
where $F_{\mu\nu}$ is field strength of photon, with $\tilde F_{\mu\nu}\equiv(1/2)\epsilon_{\mu\nu\rho\sigma}F^{\rho\sigma}$, $ V_{\mu \nu}$ is field strength of photon and $Z$ boson, with $\tilde V_{\mu\nu}\equiv(1/2)\epsilon_{\mu\nu\rho\sigma}V^{\rho\sigma}$, and  $\theta_f^i =  \tan^{-1} \frac{\tilde c_f^i}{c_f^i}$ defines the CP-phase of the Yukawa couplings. These operators can be inserted in the Barr-Zee EDM diagrams. Integrating out the internal degrees of freedoms, we have 
\begin{eqnarray}
{\cal L}_{\rm eff}&=& -i d_e \bar e \sigma^{\mu\nu} \gamma_5 e \partial_\mu A_\nu \ ,
\end{eqnarray}
with its contribution to the eEDM given by
\bea
\frac{d_e}{e}&=&\frac{\alpha m_e}{4 \pi^3 v^2} \sum_i \nonumber \\ &&  \left[-c_e^i \tilde c_\gamma^i \log\left(\frac{\tilde \Lambda^{i2}_{\rm UV}}{m_{h_i}^2}\right)+ \tilde c_e^i c_\gamma^i \log\left(\frac{\Lambda^{i2}_{\rm UV}}{m_{h_i}^2}\right)\right].  \label{eEDM}
\eea
Here $v=$246 GeV is the normalized vacuum expectation value (VEV) of the Higgs fields, and $\Lambda_{\rm UV}^i$ ($\tilde\Lambda_{\rm UV}^i$) is the relevant scale for the $h_iF^{\mu\nu}V_{\mu\nu}$ ($h_iF^{\mu\nu}\tilde V_{\mu\nu}$) operator. It is clear that the Barr-Zee contributions depend on not only the CP-odd Higgs di-photon coupling $\tilde c_\gamma^i$, but also the CP-even one $c_\gamma^i$ if the Higgs bosons have a  CP-odd coupling with electrons ($\tilde c_e^i \neq 0$). 

\begin{figure}[h]
\includegraphics[width=.95\columnwidth]{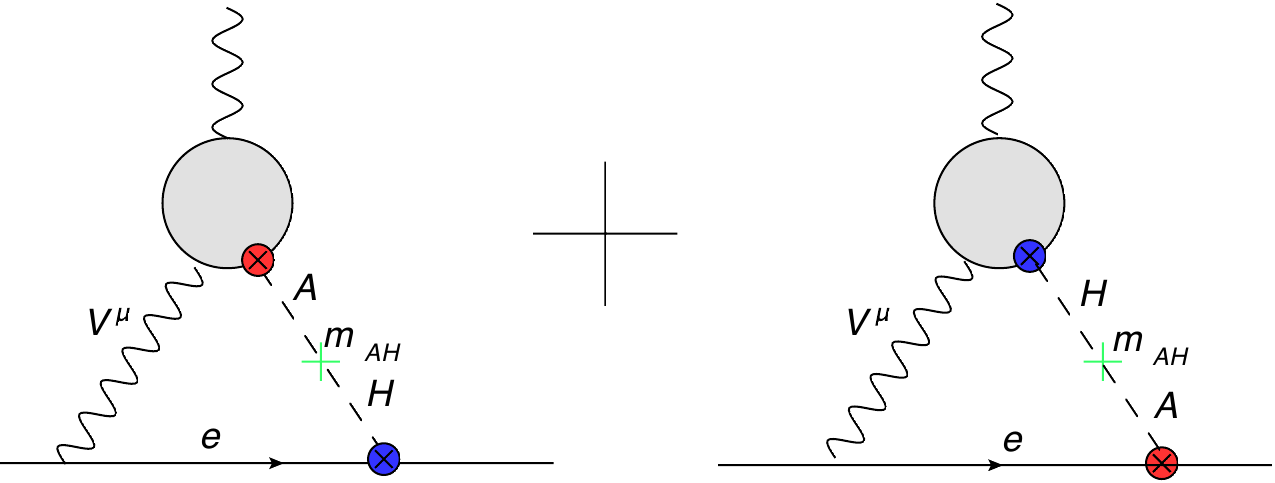} \\
\includegraphics[width=.95\columnwidth]{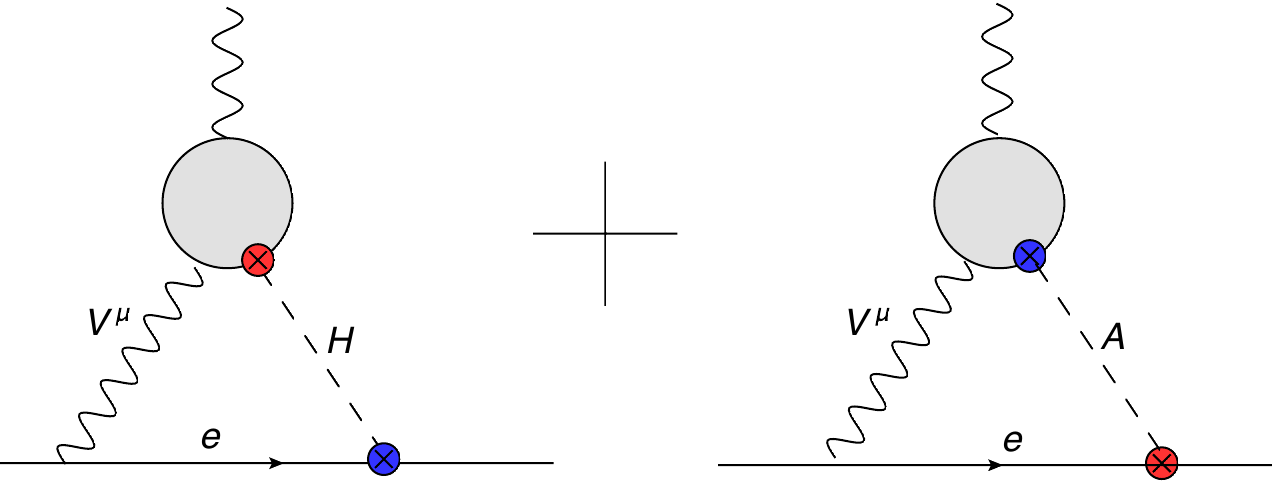}
\caption{Two cancellation mechanisms of the Bar-Zee contributions to the eEDM. Upper: cancellation occurs between the contributions of a CP-mixing Higgs boson. Bottom: cancellation occurs between the contributions of a CP-even and a CP-odd Higgs bosons.}\label{zee}
\end{figure}

The ACME measurement greatly improved the current bound on the eEDM, leading to 
\bea
\sum_i   \left[-c_e^i \tilde c_\gamma^i \log\left(\frac{\tilde \Lambda^{i2}_{\rm UV}}{m_{h_i}^2}\right)+ \tilde c_e^i c_\gamma^i \log\left(\frac{\Lambda^{i2}_{\rm UV}}{m_{h_i}^2}\right)\right]<0.14
\eea
This strongly constrains the allowed CP-violation in a single Higgs coupling, e.g., in the case with one (SM-like) Higgs only and meanwhile $\tilde c_e = 0$ \cite{McKeen:2012av}, unless this Higgs boson is decoupled. However, if a cancellation occurs among these inference terms, CP symmetry is allowed to be significantly violated, without contradicting with the current eEDM bound. Below we will use the type II 2HDM and the MSSM to show two different cancellation mechanisms, both of which are mainly motivated by EWBG: (1) cancellation occurs between the contributions of a CP-mixing Higgs boson, while the other Higgs bosons are decoupled (see the upper diagrams in Fig.~\ref{zee}); and (2) cancellation occurs between the contributions of a CP-even and a CP-odd Higgs bosons (see the bottom diagrams in Fig.~\ref{zee}).

\section{Type II 2HDM}

As an illustration, we consider type II 2HDM with a softly $Z_2$ symmetry ($\phi_1 \rightarrow -\phi_1$ and $\phi_2 \rightarrow \phi_2$) \cite{Glashow:1976nt}. 
Its tree-level Higgs potential is given by:
\begin{eqnarray}                    \label{Eq:gko-pot}
\!\!V\!\!&=&\!\!\frac{\lambda_1}{2}(\phi_1^\dagger\phi_1)^2
+\frac{\lambda_2}{2}(\phi_2^\dagger\phi_2)^2
+\lambda_3(\phi_1^\dagger\phi_1) (\phi_2^\dagger\phi_2)  \\
\!\!&+&\!\!\lambda_4(\phi_1^\dagger\phi_2) (\phi_2^\dagger\phi_1) +\frac{1}{2}\left[\lambda_5(\phi_1^\dagger\phi_2)^2+{\rm h.c.}\right] \nonumber \\
\!\!&-&\!\!\frac{1}{2}\left\{m_{11}^2(\phi_1^\dagger\phi_1)
+\!\left[m_{12}^2 (\phi_1^\dagger\phi_2)+{\rm h.c.}\right]\!
+m_{22}^2(\phi_2^\dagger\phi_2)\right\}.\!\!\!\!\! \nonumber 
\end{eqnarray}
Here $m_{12}$ and $\lambda_5$ are complex parameters. 
Their relative phase Arg$(\lambda_5 m_{12}^{4*} )$ leads to CP violation in the Higgs sector. 
We take the convention that both Higgs doublets $\phi_{1,2}$ carry a hypercharge of one unit 
and that the general Higgs VEVs are 
\begin{equation}
\langle \phi_1 \rangle = \left( \begin{array}{c} 0 \\ v_1 \end{array} \right), \ \
\langle \phi_2 \rangle = \left( \begin{array}{c} 0 \\ v_2 e^{i\xi} \end{array} \right) \ ,
\end{equation}
with $\sin^2\beta=|v_2|^2/(|v_1|^2+|v_2|^2)$, $v_1=v\cos\beta/\sqrt{2}$ and $|v_2|=v\sin\beta/\sqrt{2}$.
Then the unitary matrix $R$, defined to diagonalize the Higgs mass matrix {\cal M}
\begin{equation}
R{\cal M}R^{\rm T}={\rm diag}(M_{h_1}^2,M_{h_2}^2,M_{h_3}^2),
\end{equation}
in the mass engenstate $(h_1, h_2, h_3)$ can be easily figured out, given by
\begin{eqnarray}     \label{Eq:R-angles}
R=\begin{pmatrix}
-s_{\alpha}c_{\alpha_b} & c_{\alpha}c_{\alpha_b} & s_{\alpha_b} \\
s_{\alpha}s_{\alpha_b}s_{\alpha_c} - c_{\alpha}c_{\alpha_c} & -s_{\alpha}c_{\alpha_c} - c_{\alpha}s_{\alpha_b}s_{\alpha_c} & c_{\alpha_b}s_{\alpha_c} \\
s_{\alpha}s_{\alpha_b}c_{\alpha_c} + c_{\alpha}s_{\alpha_c} & s_{\alpha}s_{\alpha_c} - c_{\alpha}s_{\alpha_b}c_{\alpha_c} & c_{\alpha_b}c_{\alpha_c}
\end{pmatrix}
\end{eqnarray}
with $c_i=\cos\alpha_i$, $s_i=\sin\alpha_i$. Here $\alpha$, $\alpha_b$, $\alpha_c$ are mixing angles between two CP-even Higgs, the light CP-even and the CP-odd Higgs, and the heavy CP-even and the CP-odd Higgs, respectively. The angular range, beyond which $R$ is repeated, can be chosen as
$0<\alpha \le\pi$, $-\pi<\alpha_b\le\pi$ and $-\pi/2<\alpha_c\le\pi/2$.

The tree-level $h_1$ couplings rescaled by the SM values are given by   
\begin{eqnarray}\label{eq:cpara}
c_t&=&\frac{\cos\alpha\cos\alpha_b}{\sin\beta},\quad\quad c_b = c_e=-\frac{\sin\alpha\cos\alpha_b}{\cos\beta},\nonumber\\
\tilde{c}_t&=&-\cot\beta\sin\alpha_b,\quad \quad \tilde{c}_b = \tilde{c}_e=-\tan\beta\sin\alpha_b \ , \nonumber \\
a_V&=& \cos\alpha_b\sin(\beta-\alpha)
\end{eqnarray}
here $h_1$ is SM-like and $a_V$ represents the $h_1WW$ and $h_1ZZ$ couplings. The CP-phase of the top Yukawa coupling $\theta_t$ is given by 
\begin{eqnarray}
\tan \theta_t  = -  \frac{\cos \beta}{\cos \alpha }  \tan \alpha_b   \ . 
\end{eqnarray}
These tree-level effective coupling further contribute to $c_\gamma$ and  $\tilde c_\gamma$ at loop level 
\begin{eqnarray}
c_\gamma^t &=&  Q_f^2 c_f /2 = 2 c_t /9 \nonumber \\
\tilde{c}_\gamma^t &=& - 3 Q_f^2 \tilde{c}_f /4 = - \tilde{c}_t /3 \nonumber \\
c_\gamma^W &=& -7a_V /8 \label{para2} \ .
\end{eqnarray}
Note that the signs of $c_\gamma^i$ or $\tilde{c}_\gamma^i$ are derived in the convention of Eq. (\ref{eq:EFT}). To achieve the cancellation indicated by Eq.~(\ref{eEDM}), we have $c_\gamma^{t, W}  \tilde c_e / \tilde c_\gamma^t  c_e  \sim \tan \beta \sim 1$.

In this setup, the lightest Higgs boson $h_1$ leads to a leading-order contribution to the eEDM~\cite{Gunion:1990iv, Chemtob:1991vv, Hayashi:1994xf},
\begin{eqnarray}
\hspace{-0.5cm}
\left[\frac{d_e}{e} \right]^{h_1\gamma\gamma}_{t} &=& \frac{- 16\sqrt{2}\alpha G_F m_e}{3(4\pi)^3} \left( f\left( z_t \right)  \tilde{c}_e c_t + g\left( z_t \right) \tilde{c}_t c_e \rule{0mm}{3.5mm}\right), \label{edmt} \nonumber \\
\hspace{-0.5cm}
\left[\frac{d_e}{e}\right]^{h_1\gamma\gamma}_{W}
&=&-\frac{2\sqrt{2}\alpha G_F m_e}{(4\pi)^3}\left[ 3 f\left( z_W \right) + 5 g\left( z_W \right)\right] a_V \tilde{c}_e\label{edmw}
\end{eqnarray}
via the Barr-Zee diagrams~\cite{Barr:1990vd}, where $z_t=m_t^2/m_{h_1}^2$, $z_W=m_W^2/m_{h_1}^2$ and the loop functions $f(z)$ and $g(z)$ are given in~\cite{Barr:1990vd}. Numerically, we have $f(z_t)=1.0$ and $g(z_t)=1.4$.
These quantities depend on three free parameters $\alpha$, $\alpha_b$ and $\beta$. For simplicity we will work in the alignment limit $\beta = \alpha + \pi/2$, where the free parameters are reduced to $\beta$, $\alpha_b$, with $\tan\theta_t = - \cot \beta \tan \alpha_b$, and the 125 GeV Higgs boson is SM-like, if there is no CP-violation.  The overall contribution to the eEDM is then
\begin{eqnarray}
\left[\frac{d_e}{e}\right]^{h_1\gamma\gamma}
&=&\frac{2\sqrt{2}\alpha G_F m_e}{(4\pi)^3}\left[ f' \left( z_t, \tan \beta \right) - g'\left( z_W, \tan \beta  \right)\right] \nonumber \\
& &
\sin \alpha_b \cos \alpha_b, \label{edmw}
\end{eqnarray}  
with $f'(z, x) = -8(x f(z) + g(z)/x)/3)$ and  $g'(z,x) = (3f(z)+5g(z))x$. The contributions from neutral Higgs with the $Z$ gauge boson and charged Higgs with $W$ gauge boson as the propagator are generally smaller, so we  neglect them in the calculation \cite{Chang:1990sf}.

\begin{table}[h]
\begin{tabular}{c|c|c|c}
\hline
&  $\gamma\gamma$ & $WW^*$ & $ZZ^*$ \\
\hline
ATLAS & $1.17 \pm0.27$ \cite{Aad:2014eha} & $0.99 ^{+0.31}_{-0.28}$ \cite{Aad:2013wqa} & $1.44^{+0.40}_{-0.33}$ \cite{Aad:2014eva} \\
\hline
CMS &  $1.14^{+0.26}_{-0.23}$ \cite{Khachatryan:2014ira} & $0.72^{+0.20}_{-0.18}$ \cite{Chatrchyan:2013iaa} & $0.93^{+0.29}_{-0.25}$ \cite{Chatrchyan:2013mxa} \\
\hline
& $bb$ & $\tau\tau$ &
 \\
\hline
ATLAS & $0.52\pm0.40$ \cite{Aad:2014xzb} & $1.4^{+0.5}_{-0.4}$ \cite{ATALAStautau} & \\ 
\hline
CMS & $1.15\pm0.62$ \cite{CMSbb} & $0.78\pm0.27$ \cite{Chatrchyan:2014nva} &
\\
\hline
\end{tabular}
\caption{The LHC data used for the fitting.}\label{data}
\label{cuts}
\end{table}

\begin{figure}[h]
\centerline{
\includegraphics[width=.9\columnwidth]{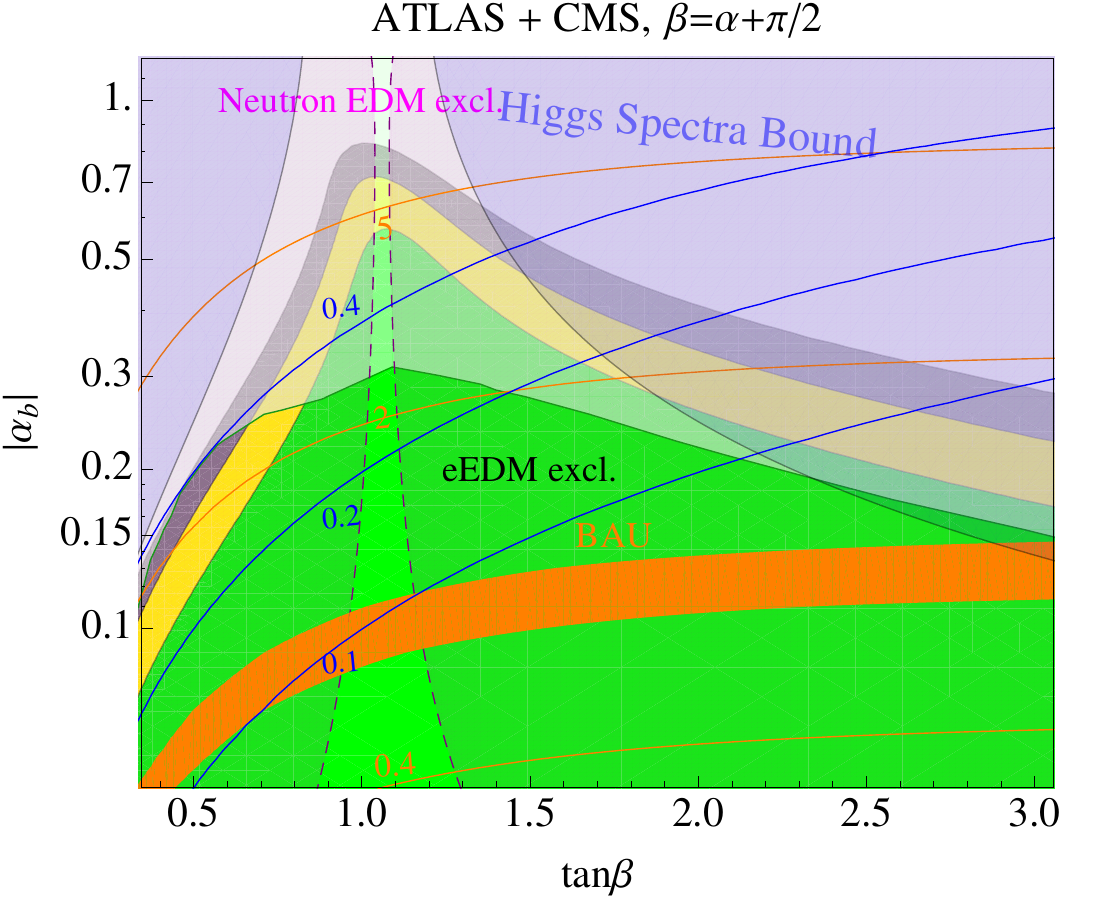}}
\caption{The allowed parameter region $|\alpha_b|$ versus $\tan \beta$ from the Higgs global fits and EDM experiments. The grey, yellow, green region are the 1,2,3 $\sigma$ allowed region for the LHC Higgs fits. The left and right region bounded by the black dashed lines and the upper region bounded by the green solid line are excluded by the eEDM and neutron EDM experiment at 90\% C.L. The constraint from Mercury experiment \cite{Griffith:2009zz} is weaker than the one from neutron EDM so we do not plot it. The light blue region is the theoretical bound from the Higgs mass spectra $m_{h}=125$ GeV, $m_{H^+}$=420 GeV, $m_{H2}=400$ GeV and $m_{H3}=450$ GeV \cite{Inoue:2014nva}. The blue solid contours represent the CP-phase of the top Yukawa coupling $\tan \theta_t$. The orange band and contours give the baryon-to-entropy density ratio in unit of $10^{-11}$. To calculate the BAU, a wall velocity $v_w=0.02$, a wall width $L_w=5/T_c$ and a critical temperature $T_c=100$ GeV are assumed  for the bubbles generated during the EWPT. 
}\label{fig:alphab}. 
\end{figure}

The fitting results are presented in Fig. \ref{fig:alphab} where the inclusive LHC data published in March 2013 (see Table~\ref{data}) and the most recent ACME results~\cite{Baron:2013eja} are applied. In the presence of CP-even and CP-odd Higgs mixing, both the current ACME constraints and the Higgs global fits favors the region with $\tan \beta \sim 1$, where we have $\tan \theta_t \sim - \tan \alpha_b$. The former is easy to understand since a cancellation between $f'(z_t, \tan \beta)$  and $g'(z_W, \tan \beta)$ in Eq. (\ref{edmw}) requires $\tan \beta = 1.04$. The latter is because a relatively small $\tan\beta$ can help avoid a too large signal rate of $h\to bb$, and hence an over-suppressed $h\to \gamma\gamma$ rate (see Table~\ref{data}). In this cancellation region, a CP-violation effect with $ |\tan \alpha_b|, |\tan \theta_t| >  0.1$ is allowed while the most stringent constraints are from the nEDM.

\section{MSSM}

Though the MSSM is of type II 2HDM, there is no tree-level CP-violation in the Higgs sector either explicitly or spontaneously~\footnote{This is different from what happens in the singlet-extensions of the MSSM where tree-level CP-violation is possible in the Higgs sector.}, due to a vanishing $\lambda_5$ term in Eq.(\ref{Eq:gko-pot}). 
So, the CP-phases used for EWBG mostly arise in the tree-level superparticle sectors, such as the chargino, neutrolino, squark and slepton sectors. 

The explicit CP-violation in these sectors can break the CP symmetry in the Higgs sector at loop level, leading to CP-even and CP-odd mixing terms in the Higgs squared mass matrix. But the Higgs CP-mixture caused by this effect is small due to the loop suppression. For nonstandard Higgs bosons we notice that the CP-mixture is typically below 10\%, consistent with~\cite{Arbey:2014msa}, even if the CP-phase arises from the stop sector, while for the SM-like Higgs boson, the CP-mixture is suppressed more by an extra $\tan\beta$ factor.   So the Higgs eigenstate are approximately CP-eigenstates, with their couplings with electrons are either $|c_e| \gg |\tilde c_e|$ (for CP-even Higgs bosons) or $|c_e| \ll |\tilde c_e|$ (for CP-odd Higgs bosons). On the other hand, a relatively large $\tan\beta$ is favored in the MSSM, given that the tree-level mass of the SM-like Higgs boson is larger in this case. This leads to $|c_e^h| \ll |c_e^{H,A}|$, and hence a small $h$ contribution to the eEDM (mainly via the $c_e \tilde c_\gamma$ term). So in the MSSM with the EWBG mechanism implemented,  the main contributions to the eEDM are made by nonstandard Higgs bosons unless they are highly decoupled.          

Among these CP-violating sources, the one arising in the chargino sector is of particular interest because of its high efficiency in generating the BAU via EWBG. The charginos enter the $H\gamma\gamma$ and the $A\gamma\gamma$ loops as new mediators, inducing non-trivial contributions to the eEDM via the $c_e\tilde c_\gamma$ and $\tilde c_e c_\gamma$ terms, respectively. Though these two contributions are comparable in magnitude, their signs are typically different. Given an extra minus sign for the term $c_e^i \tilde c_\gamma^i$ in Eq.(\ref{eEDM}), this scenario is strongly constrained by the ACME eEDM bound \cite{Brod:2013cka}. However, recall that charged particles like stau leptons enter the $A\gamma\gamma$ loop as well.  If a non-trivial CP-phase is turned on in the stau sector, new contributions to the eEDM will be introduced via the $\tilde c_e c_\gamma$  term. (Note, such a CP-violating coupling will not induce non-trivial contributions to the eEDM via the $H\gamma\gamma$ loop or the $c_e \tilde c_\gamma$  term, since stau leptons are scalar particles.) This provides a potential cancellation, such that a CP-phase in the Higgs-chargino couplings which is large enough for implementing the EWBG mechanism, is still allowed. 

With the CP-violation turned on in the stau sector, the main contributions to the eEDM are then given by 
\begin{eqnarray}\label{MSSMedm}
\left[\frac{de}{e}\right]  &\approx& 
C \tilde{c}^A_e  \sum_{j=1,2}   \left( c^{\tilde{\chi}^\pm_{j}}_\gamma \ln \frac{1}{z_{\tilde{\chi}^\pm_j}^A} + c^{\tilde{\tau}^\pm_{j}}_\gamma  \ln\frac{1}{z_{\tilde{\tau}^\pm_j}^A}  \right)
  \nonumber \\ && 
- C c^H_e \sum_{j=1,2} \tilde{c}^{\tilde{\chi}^\pm_{j}}_\gamma  \ln\frac{1}{z_{\tilde{\chi}^\pm_j}^H}
\end{eqnarray}
with the terms in the first and the second lines mediated by $A$ and $H$, respectively. Here 
$C= \frac{ \alpha\,m_e}{4\pi^3 v^2}$, $z_x^y=  \frac{m_x^2}{m^2_y} $ and 
\begin{eqnarray}
 c_\gamma^{\tilde{\chi}_i^\pm}=-\sum_{i}
\frac{M_W}{2\sqrt{2}\;m_{\tilde{\chi}_i^\pm}} \tilde{c}_{\tilde{\chi}_i^\pm},  c_\gamma^{\tilde{\tau}_i}=-\sum_{i} 
\frac{M_W}{2\sqrt{2}\;m_{\tilde{\tau}_i}} \tilde{c}_{\tilde{\tau}_i}
\end{eqnarray}
As for $m_{\tilde{\chi}_i^\pm}$, $\tilde{c}_{\tilde{\chi}_i^\pm}$(CP-odd interaction between $H$ and charginos), we refer to \cite{Lee:2003nta} for the details. 

\begin{figure}[!htp]
  \includegraphics[width=0.8\linewidth]{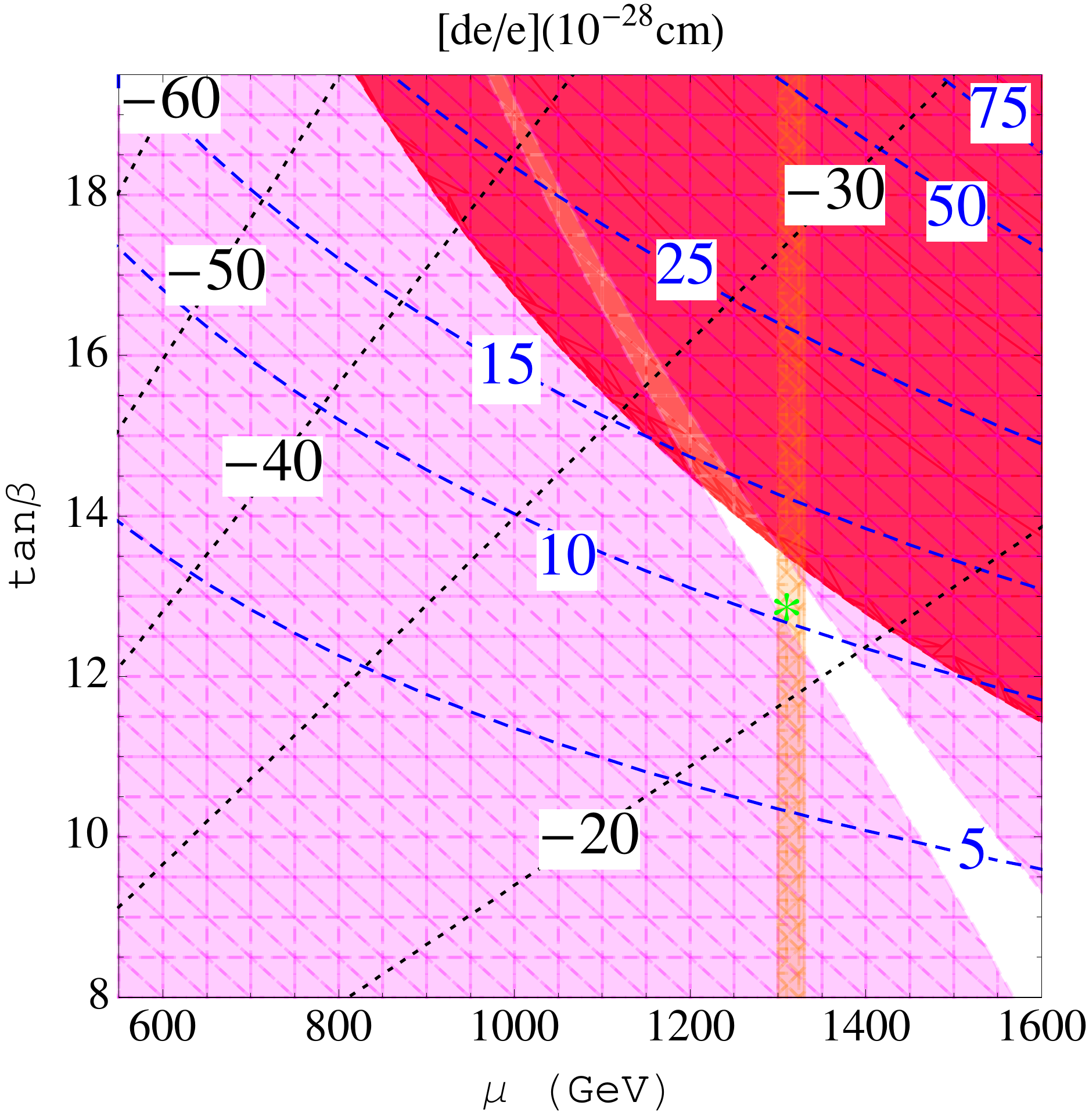}
  \caption{
Parameter region allowed by the current eEDM, mercury EDM and neutron EDM bounds, and favored by the EWBG (orange band) in the $\mu-\tan \beta$ plane.  The chargino, stau contributions to the eEDM are plotted with black dashed and blue dashed contours, respectively. The magenta and red regions have been excluded by the ACME and the Mercury experiments~\cite{Griffith:2009zz}. The bounds of neutron EDM are much weaker and fall outside of the figure. Here we are assuming: charged Higgs mass $m_{H^\pm}=450$ GeV; trilinear softly SUSY-breaking parameters (except $A_t$) 1.5 TeV and $A_t$ = 2.5 TeV; soft masses of gauginos $M_1=0.1$ TeV, $M_2=0.45$ TeV and $M_3=3.5$ TeV; soft masses of squarks and sleptons $0.1M_{Q_1,U_1,D_1}= 0.1M_{Q_2,U_2,D_2}= M_{Q_3,U_3,D_3}= 1.2$ TeV,  and $0.1M_{L_1}=0.1M_{L_2}=M_{L_3}=0.15$ TeV, $0.05M_{E_1}=0.05M_{E_2}=M_{E_3}=0.35$ TeV, and CP-phases  ${\rm Arg}(\mu M_2^*)={\rm Arg}(\mu A_f^*)=90^{o}$. With their mass set to be $\sim 10$ TeV, the contributions of the squarks and sleptons of the first two generations to the one-loop EDM are suppressed. To calculate the baryon asymmetry generated via EWBG, we closely follow the work done in~\cite{Lee:2004we}, with an expansion velocity $v_w=0.02$, a wall width $L_w=5/T_c$ and a critical temperature $T_c=100$ GeV assumed  for the bubbles generated during the EWPT. 
}\label{fig:edmmuAtbau1}
\end{figure}

Fig.~\ref{fig:edmmuAtbau1} depicts all low energy experimental constraints
calculated by the ${\tt CPsuperH}$~\cite{Lee:2003nta, Lee:2012wa}. Note, 
however, the ${\tt CPsuperH}$ codes used in the analysis was revised by the authors or this letter (for details, see Appendix C), 
in which we corrected a sign error in the anomalous dimension of the dipole operators, 
updated the QCD hadron matrix elements, and incorporated the mixing effects between 
the dipole and psedo-scalar operators during renormalization group running, and the missed contributions to the eEDM of the $W$-mediated $h\gamma\gamma$ loop in the Barr-Zee diagrams.
With these corrections, we notice that the neutron and mercury bounds may become substantially weaker than the ones given by the original CPsuperH.

As indicated in Fig.~\ref{fig:edmmuAtbau1}, the charginos have a negative contribution to the eEDM (black-dashed contours), by coupling with both $A$ and $H$. The staus, on the other hand, have a positive contribution to the eEDM (blue-dashed contours), by mainly coupling with $A$. Both contributions are enhanced by $\tan \beta$ because of $\tilde{c}^A_e \propto \tan \beta $ and ${c}^H_e \propto 1/ \cos \beta $. Their dependences on the $\mu$ parameter however is different. For charginos, the $H/A$-chragino-chargino couplings $g_ {H \chi^\pm_i \chi^\pm_i}^P \approx i (C_R)_ {i1}  (C_L)^*_ {i2}/2 - h.c.$ and $g_ {A \chi^\pm_i \chi^\pm_i}^S \approx - i (C_R)_ {i1}  (C_L)^*_ {i2}/2 - h.c.$, $i=1,2$ \cite{Ellis:2008zy}. With $\mu > M_2$ assumed here, a small $\mu$ value will increase the off-diagonal term in the chargino mixing matrices, $(C_L)_ {12}$ and $(C_R)_ {21}$, and hence the overall eEDM contribution. As for staus, a larger $\mu$ leads to a lighter $\tilde \tau_1$ because of a larger mixture term $\propto |\mu \tan \beta -A_\tau|$. Therefore, their contribution to the eEDM increases for a larger $\mu$ value. Due to their cancellation there exists a blank region in Fig.~\ref{fig:edmmuAtbau1} where the total eEDM is below the current ACME bound. This region overlaps with the EWBG favored region which has been excluded by the ACME bound, if only the chargino contribution is taken into account.  One benchmark point is presented in Table ~\ref{tab:BP1}.

 \vspace*{0.5cm}
\begin{table}[!h]
\centering
\begin{tabular}{cccccccccccccccccccc}
\hline
 $R_{\gamma\gamma}$ &$\tan \beta$&$\mu$(TeV) &$m_{H_1}$(GeV)& 
&\\
\hline
   0.84&$ 12.8 $&1.31&125.4& \\ [+1mm]
\hline
 [de/e] (cm)& $[d_{Hg}/e]$(cm)&$[d_n/e]$(cm)&$n_b/s$
 &\\
 \hline
 4.2$\times 10^{-29}$&3.0$\times 10^{-29}$&9.6$\times 10^{-27}$&0.85$\times 10^{-10}$&\\[+1mm]
\hline
\end{tabular}
\def\baselinestretch{1.1}
\caption{An benchmark in the MSSM, with the other parameters set as Fig.\ref{fig:edmmuAtbau1}.}
\def\baselinestretch{1.0}
\label{tab:BP1}
\end{table}
 
\section{Discussion and Conclusion}

In this letter, we present a class of cancellation mechanisms to suppress the total contributions of Barr-Zee diagrams to the eEDM, which may occur either between the contributions of a CP-mixing Higgs boson, with the other Higgs bosons being decoupled, or between the contributions of a CP-even and a CP-odd Higgs bosons. As an illustration, we study two scenarios: the type-II 2HDM where the tree-level CP-phase arises from the Higgs sector,  and the MSSM where the tree-level CP-phase arises from Higgs-superparticle interaction sectors. In the 2HDM, $\tan\beta \sim 1$ is favored by the LHC Higgs bounds and the contributions of the Barr-Zee diagrams to the eEDM are mainly mediated by a CP-mixed SM-like Higgs boson. With a cancellation between them, a CP-phase as large as $\mathcal O(0.1-1)$ is still allowed for the top Yukawa coupling, induced by the Higgs CP mixing. In the MSSM, a large $\tan\beta$ is favored by the LHC Higgs bounds, and the contributions of the Barr-Zee diagrams to the eEDM are mainly mediated by nonstandard CP-even and CP-odd neutral Higgs bosons (if they are not as heavy as 10 TeV scale or above). With a cancellation between them,  a maximal CP-phase in their couplings with superparticles like charginos is still allowed. In both cases a successful EWBG is possible, which turns out to be challenging without the assistance of these cancellation mechanisms. 

The EWPT, another key element for the EWBG, is not addressed in this letter, since it is ``orthogonal'' to the discussions above. It is straightforward to generalize the discussions on the eEDM cancellation to the 
2HDM + a singlet, or the MSSM extensions with an extra singlet superfield or gauge group, where  a strong enough EWPT is not difficult to achieve (e.g., see \cite{Huang:2014ifa, Kozaczuk:2014kva, Kang:2004pp}. Furthermore, in the MSSM while an extremely light right handed stop scenario ($\sim 110$ GeV) \cite{Carena:1996wj, Delepine:1996vn} has been ruled out by the Higgs global fits \cite{Carena:2012np, Cohen:2012zza, Curtin:2012aa} and direct searches, a moderately light stop with another light scalar particles (like sbotoom or stau) may still be viable \cite{Huang:2012wn}. The detailed discussions in this regard will be left for future studies.


\section*{Acknowledgments}

We would like to thank Michael Ramsey-Musuolf, Carlos Wagner and Yue Zhang for useful discussions; and David McKeen for collaboration at the early stages
of this project. TL is supported by his start-up fund at the Hong Kong University of Science and Technology.
We also would like to acknowledge the hospitality of the Kavli Institute for Theoretical Physics and the Aspen Center for Physics (Simons Foundation), where part of this work was completed.

\appendix

\section*{Neutron and Mercury EDMs in the MSSM:  \\ the {\tt CPsuperH} vs. Our Analysis}
\label{sec:CPsuperH}

In the Appendix, we will introduce the modifications and corrections made to the  {\tt CPsuperH} in our analysis. Then we will make a comparison between the constraints of neutron and Mercury EDMs given by the original {\tt CPsuperH} and the modified one. As a start, let's simply introduce the theoretical methods used for calculating the neutron and Mercury EDMs, basically following ref~\cite{Engel:2013lsa,Dekens:2013zca}.\\

{\bf (1) Renormalization Group Running of Wilson Coefficients} \\

The  EDM of a fermion ($d^E_f$; electron EDM is an exception which is denoted by $d_e$ in this letter), the Chromo-EDM  of a quark ($d^C_q$), and the Weinberg operator coupling ($d^G$)
are defined by the Lagrangian 
\begin{eqnarray}
 \label{CEDM}
{\cal L}_{\rm
  (C)EDM} &=&
  -\frac{i}{2}\,d^E_f\,F^{\mu\nu}\,\bar{f}\,\sigma_{\mu\nu}\gamma_5\,f   \nonumber \\
&&- \frac{i}{2}\,d_q^C\,  g_s G^{a\,\mu\nu}\,\bar{q}\,\sigma_{\mu\nu}\gamma_5 T^a q  \nonumber \\
&& {+}\frac{1}{6}\,d^{\,G}\,f^{abc}\,\epsilon^{\mu\nu\lambda\sigma}G^a_{\rho\mu}
G^b_{\lambda\sigma}
{G^c}_{\nu}^{~~\rho}\; . 
\end{eqnarray}
Here $F^{\mu\nu}$ and $G^{a\,\mu\nu}$ are  the  electromagnetic and
strong  field strengths, $T^a=\lambda^a/2$ denotes the
generators of  the  SU(3)$_C$ group. In  the MSSM, the
Weinberg operator $d^{\,G}$ is given by:
\begin{equation}
  \label{dG}
d^{\,G}\ =\ (d^{\,G})^{\tilde{g}}\: +\: (d^{\,G})^{H}\; ,
\end{equation}
with $(d^{\,G})^{\tilde{g}}$ being contributed by the quark-squark-gluino coupling~\cite{Dai:1990xh} and  $(d^{\,G})^{H}$ being contributed by neutral Higgs bosons~\cite{Weinberg:1989dx,Dicus:1989va}.

Defining the wilson coefficients as 
\begin{eqnarray}
{\delta_f \equiv -\frac{\Lambda^2 d_f^E}{2 e Q_f m_f}, }\ \ \  \tilde \delta_q \equiv -\frac{\Lambda^2 d_q^C}{2 m_q}, \ \ \   C_{\tilde G} = \frac{\Lambda^2
d^G}{3 g_s}\ \ \  ,
\end{eqnarray}
with $m_f$ and  $\Lambda$ denoting the fermion masses and the MSSM scale respectively,
the Lagrangian (\ref{CEDM}) can be rewritten as 
\begin{eqnarray}
{\cal L}_{(C)EDM}&=&i\sum_f\, \frac{\delta_f}{\Lambda^2} m_{f} Q_f e F^{\mu\nu} \bar f \sigma_{\mu\nu} \gamma_5 f  \nonumber\\
&&+ i \sum_q\, \frac{\tilde{\delta}_q}{\Lambda^2} m_{q} g_s G^{a\mu\nu} \bar q \sigma_{\mu\nu} \gamma_5 T^a q \nonumber\\
&&+ \frac{C_{\tilde{G}}}{2 \Lambda^2} g_s f^{abc} \epsilon^{\mu\nu\lambda\sigma}G^a_{\rho\mu}\,
G^b_{\lambda\sigma}
{G^c}_{\nu}^{~~\rho}\; .
\end{eqnarray}

To calculate the neutron and mercury EDMs, we need to incorporate the effect of renormalization group running of the WIlson coefficients 
from new physics scale to a hadron scale. During this process, flavor-conserving CP-odd four-fermion operators may lead to nontrivial corrections 
to the Wilson coefficients of the CEDM and Weinberg operators via mixing. So a more complete Lagrangian for the calculation of the neutron 
and mercury EDMs should be 
\begin{eqnarray}
{\cal L}_\text{CPV} &=& {\cal L}_{(C)EDM}+
 \sum_q \frac{C_4^q}{\Lambda^2} {\mathcal O}_4^q
\nonumber \\&& +\sum_{q'\ne q} \frac{\widetilde{C}_1^{q'q} }{\Lambda^2}\widetilde{\mathcal O}_1^{q'q}
+ \frac{1}{2}\sum_{q'\ne q}\frac{\widetilde{C}_4^{q'q}}{\Lambda^2}\widetilde{\mathcal O}_4^{q'q} \ .
\label{effop}
\end{eqnarray}
Here the first two CP-odd four-fermion operators 
\begin{eqnarray}
{\mathcal O}_4^q
&=&\, \overline{q} q \overline{q} \,i\gamma_5 q\,, \nonumber\\
%
\widetilde{\mathcal O}_1^{q'q}
&=&\, \overline{q'} q' \overline{q} \,i\gamma_5 q 
\label{4foperator}
\end{eqnarray}
can be generated through
CP-violating  neutral Higgs-boson  mixing in the  $t$-channel and
CP-violating  Yukawa threshold  corrections. The corresponding CP-odd coefficients are given by 
\begin{eqnarray}
  \label{eq:cff}  
C_4^q\ &=&\ g_q\, g_{q}\,
\frac{c_{q}\,\tilde c_{q}}{M_{H}^2}\; ,\nonumber\\
\widetilde{C}_1^{q'q}\ &=&\ g_{q'}\, g_q\,
\frac{c_{q'}\,\tilde c_{q}}{M_{H}^2}\; ,
\end{eqnarray}
with  $g_{q(q')}=m_{q(q')}/v$  and  $v=2M_W/g$. 
The last CP-odd four-fermion operator
\bea
\widetilde{\mathcal O}_4^{q'q}
&=&\overline{q'_\alpha} \sigma^{\mu\nu} q'_\beta \overline{q_\beta} \,i\sigma_{\mu\nu}\gamma_5 q_\alpha\,,
\label{add4foperator}
\eea
on the other hand, is generated from the operatror mixing effects of $\widetilde{C}_1^{q'q}$ and $\widetilde{C}_1^{qq'}$ which follow the Eq.~(\ref{Eq:Gamma}) below.
To calculate the Wilson coefficients $\left(\frac{\delta_q}{Q_q}, \tilde{\delta}_q, - \frac{3 C_{\tilde{G}}}{2} \right)$ at a GeV scale,
we need to take an evolution for 
${\bf C}=\left(\frac{\delta_q}{Q_q}, \tilde{\delta}_q, - \frac{3 C_{\tilde{G}}}{2},C_4^q,\widetilde{C}_1^{q'q},\widetilde{C}_1^{qq'},\widetilde{C}_4^{q'q}\right)$
 from the MSSM scale  $\Lambda$ down to GeV scale, based on the Renormalization Group Equations (RGE)~\cite{Degrassi:2005zd,Hisano:2012cc,Dekens:2013zca} :
\begin{eqnarray}\label{RGE}
\frac{d}{d \ln \mu}\bold{C} =  
\bold{C}\cdot \bold{\Gamma}\,
\end{eqnarray}
The one-loop anomalous dimension matrix is given by
\begin{align}
{\bf \Gamma} = \begin{bmatrix}
\frac{\alpha_s}{4\pi} \gamma_s & {\bf 0}                        & {\bf 0} \\
\frac1{(4\pi)^2} \gamma_{sf}   & \frac{\alpha_s}{4\pi} \gamma_f & {\bf 0} \\
\frac1{(4\pi)^2} \gamma'_{sf}  & {\bf 0}                        & \frac{\alpha_s}{4\pi} \gamma'_f
\end{bmatrix}, \label{Eq:Gamma}
\end{align}
with
\begin{align}
\gamma_s &=
\begin{bmatrix}
+ 8 C_F & 0         & 0 \\
+8C_F & +16C_F-4N & 0 \\
0     & +2N       & N+2n_f+\beta_0 
\end{bmatrix}, \label{Eq:Gamma_s}
\end{align}
\begin{align}
\gamma_f &=
\begin{bmatrix}
-12C_F+6     \end{bmatrix}, \label{Eq:Gamma_f}
\end{align}
\begin{align}
\gamma'_f &=
\begin{bmatrix}
-12C_F                      & 0       & -1      \\
0                           & -12C_F   &-1\\
-12 &-12&-8C_F-\frac{6}{N}        \end{bmatrix}, \label{Eq:Gamma_f'}
\end{align}
\begin{align}
\gamma'_{sf} &=
\begin{bmatrix}
0&0&0\\
0 & 0  &0    \\
-16\frac{m_q'}{m_q}\frac{Q_q'}{Q_q}&-16\frac{m_q'}{m_q}&0
\end{bmatrix}, \label{Eq:Gamma'_sf}
\end{align}
and
\begin{align}
\gamma_{sf} &=
\begin{bmatrix}
+4 & +4 &0
\end{bmatrix}. \label{Eq:Gamma_sf}
\end{align}
Here $N=3,~C_F = (N^2-1)/(2N)=4/3,~\beta_0 = (11N-2n_f)/3$, $n_f$ is the flavor number, and  $q$ runs over $u, d, b$,.

Besides the RG running and operator mixing effects calculated from Eq.(\ref{RGE}) which give corrections to the Wilson coefficients $\delta_q$, $\tilde{\delta}_q$ and $C_{\tilde{G}}$, there are two additional bottom quark threshold effects which needs to be emphasized below. First, the Weinberg operator receives a shift from the bottom quark CEDM after the bottom quark is integrated out at $m_b$~\cite{Boyd:1990bx, Hisano:2012cc}:
\begin{eqnarray}
\Delta C_{\tilde{G}} (m_b) = \frac{\alpha_S(m_b)}{12\pi} \tilde{\delta}_b(m_b) \ . \label{3gluonshift}
\end{eqnarray}
Here $\tilde{\delta}_b(m_b)   =  \tilde{\delta}^0_b(m_b)  +  \Delta \tilde{\delta}_b(m_b)$ is  the $b$-quark CEDM at the $m_b$ scale, with $\tilde{\delta}^0_b(m_b)$
being the direct CEDM calculated from two-loop Barr-Zee graphs. The $\Delta \tilde{\delta}_b(m_b)$ is the bottom quark CEDM correction from RG running and operator mixing. 
Keeping the leading logarithmic terms that contributes to $\Delta \tilde{\delta}_b(m_b)$ at the matching scale $\mu=m_b$, we have 
\begin{eqnarray}
\Delta \tilde{\delta}_b (m_b) &\approx& \frac{1}{8\pi^2}  C_4^b (M_H) \log\frac{M_H}{m_b} \ ,\label{shiftC} 
\end{eqnarray}
which could be figured out from Eq.~(\ref{Eq:Gamma}) and is from integrating out b quark at one-loop level.

The second important corrections by integrating out the bottom quark are the shift to CEDMs of quarks:
\begin{eqnarray}
\Delta \tilde{\delta}_q(m_b) &\approx&  \frac{g_s^2}{64\pi^4} \frac{m_b}{m_q}  \nonumber \\&&  ( \tilde{C}_1^{bq} (M_H)+\tilde{ C}_1^{qb} (M_H) ) \left(\log\frac{M_H}{m_b}\right)^2  \label{edmshift}
\end{eqnarray}
which are actually a two loop effects where the operator $ \tilde{C}_1^{bq}$ and $\tilde{ C}_1^{qb}$ first mix into $\widetilde{C}_4^{qq'}$ and then mix into $\Delta \tilde{\delta}_q(m_b)$. Numerically, the suppression from one additional loop is compensated by the $m_b/m_q$ enhancement.  \\

{\bf (2) Neutron and Mercury EDMs} \\

The neutron EDM is calculated by
\begin{eqnarray}
d_n = \left( e \zeta_n^u \delta_u + e \zeta_n^d \delta_d \right) + 
\left( e \tilde \zeta_n^u \tilde\delta_u + e \tilde\zeta_n^d \tilde\delta_d \right) + \beta_n^{G} C_{\tilde{G}} \ ,
\end{eqnarray}
In the {\tt CPsuperH}, we updated hadronic matrix elements with $\zeta_n^u=0.82\times10^{-8}$, $\zeta_n^d=-3.3\times10^{-8}$, $\tilde\zeta_n^u=0.82\times10^{-8}$, $\tilde \zeta_n^d=1.63\times10^{-8}$ and $\beta_n^{G}=2 \times 10^{-20} \,e\,{\rm cm}$~\cite{Engel:2013lsa}.

Compared to the electron EDM $d_e$ and the CP-odd electron-nucleon interactions
\begin{eqnarray} 
{\cal       L}   &=& C_S \bar e i\gamma_5 e  \bar N N   \nonumber \\ &&   
+ \ C_P\,\bar{e}e\,\bar{N}i\gamma_5 N\: +\: 
C^\prime_P\,\bar{e}e\,\bar{N}i\gamma_5 \tau_3  N\; ,  \label{enint}
\end{eqnarray}  
the nuclear Schiff moment ($S$) has a larger contribution to the mercury EDM. The Schiff moment is generated by long-range, pion-exchange mediated P- and T-violating nucleon-nucleon interactions, 
\begin{eqnarray}
\label{eq:piNN1}
\mathcal{L}_{\pi NN}^\mathrm{TVPV} = {\bar N}\left[ \bar g_{\pi}^{(0)} {\vec\tau}\cdot{\vec \pi} +  \bar g_{\pi}^{(1)} \pi^0 +\bar g_{\pi}^{(2)}(2\tau_3\pi^0-{\vec\tau}\cdot{\vec \pi})\right]N .
\end{eqnarray}
In a general context, the isoscalar and isovector couplings $\bar g_{\pi}^{(0)},~\bar g_{\pi}^{(1)}$ are dominant over the isotensor coupling $\bar g_{\pi}^{(2)}$~\cite{Engel:2013lsa}, 
so the mercury EDM is approximately given by~\cite{Engel:2013lsa},
\begin{eqnarray}
d_{\rm Hg} = \kappa_S S \approx \kappa_S \frac{2 m_N g_A}{F_\pi} \left(a_0 \bar g_{\pi}^{(0)} + a_1 \bar g_{\pi}^{(1)}\right) \ ,
\end{eqnarray}
in which
\begin{eqnarray}
&&\bar g_{\pi}^{(0)} = \tilde \eta_{(0)} ({\tilde\delta}_u + {\tilde\delta}_d) + \gamma^{\tilde{G}}_{(0)} C_{\tilde{G}}, \nonumber\\
&&\bar g_{\pi}^{(1)} = \tilde \eta_{(1)} ({\tilde\delta}_u - {\tilde\delta}_d) + \gamma^{\tilde{G}}_{(1)} C_{\tilde{G}} \ ,\nonumber\\
&&g_A\approx 1.26,~F_\pi=186\, {\rm MeV}.
\end{eqnarray}
To do the calculation, in the {\tt CPsuperH} we updated hadronic matrix elements with {\cite{Engel:2013lsa}}: $\tilde \eta_{(0)}=-2\times10^{-7}$, $\tilde \eta_{(1)}=-4\times10^{-7}$, $\gamma^{\tilde{G}}_{(0)}\approx \gamma^{\tilde{G}}_{(1)} = 2\times 10^{-6}$; updated the nuclear matrix elements with \cite{oai:arXiv.org:hep-ph/0203202}: $a_0=0.01\,e\,{\rm fm}^3$, $a_1=\pm0.02\,e\,{\rm fm}^3$; and assume a new atomic sensitivity coefficient $\kappa_S=-2.8\times10^{-4}\,{\rm fm}^{-2}$ \cite{oai:arXiv.org:hep-ph/0203202}. At last we would like to emphasize again that both the sign correction of the anomalous dimension coefficient $\gamma_e$ and the mixing effect of the RGE operators  have been incorporated in calculating  $\{\delta_u, \delta_d, \tilde{\delta}_u, \tilde{\delta}_d, C_{\tilde{G}}\}$ at a hadron scale in our analysis, which may cause an important change for the bounds of both the neutron and mercury EDMs given by the original {\tt CPsuperH}.\\

{\bf(3) The {\tt CPsuperH} vs. Our Analysis}\\

\begin{figure}[!htp]
    \includegraphics[width=0.8\linewidth]{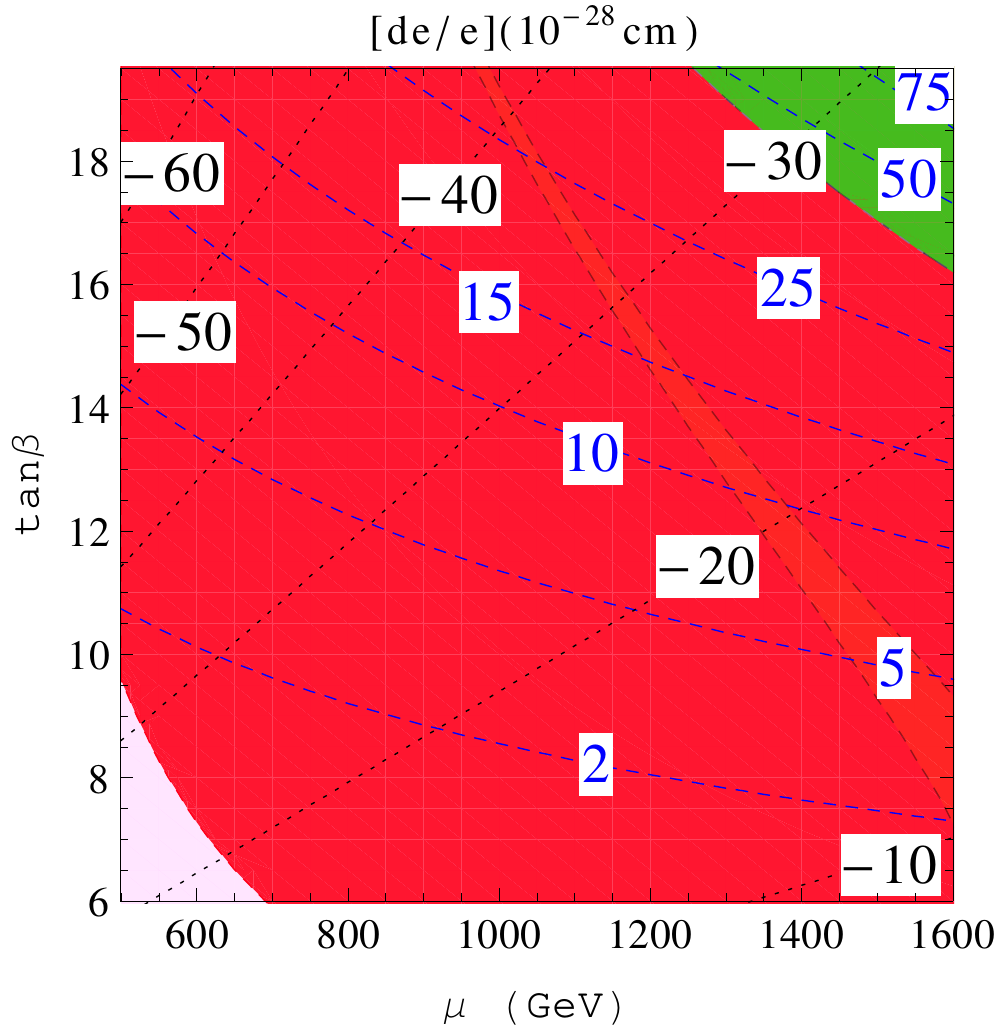}
  \caption{The neutron and mercury EDM bounds before the revision of the {\tt CPsuperH}, with the parameter values assumed to be the same as the ones defined in Fig.~\ref{fig:edmmuAtbau1}. The green and the red regions are excluded by the neutron and mercury EDM bounds, respectively. }\label{fig:cps}
\end{figure}

In the {\tt CPsuperH}, we made the following corrections and modifications: 
\begin{enumerate}

\item A sign error in an anomalous-dimension coefficient $\gamma_e$ was fixed, which is equal to the (1,1) component of Eq. (\ref{Eq:Gamma_s}), up to some numerical factor caused by different definitions. The {\tt CPsuperH} follows Ref.~\cite{Arnowitt:1990eh} where $\gamma_e = 8/3$ is defined. As pointed in Ref.~\cite{Giudice:2005rz,Degrassi:2005zd}, however, $\gamma_e$ should be $-8/3$ rather than $8/3$ in the notation of Ref.~\cite{Arnowitt:1990eh}.

\item The mixing effects between the Weinberg's gluonic operator, quark color dipole operator, quark electric dipole operator and the CP-odd four-fermion operators which are defined by Eq. (\ref{3gluonshift}-\ref{edmshift})  are incorporated in the {\tt CPsuperH}. 

\item The hadronic matrix elements are updated. 

\item The Barr-Zee diagrams with the Higgs-photon-photon($Z$ boson) loop mediated by a $W$ boson are incorporated.

\end{enumerate}
These corrections and modifications made to the {\tt CPsuperH} may significantly change the neutron and the mercury EDM bounds for CP-violation in the MSSM. As a comparison, the neutron and mercury EDM bounds after and before the revision of the {\tt CPsuperH} are shown in Fig.~ \ref{fig:edmmuAtbau1} and Fig.~\ref{fig:cps}, respectively, with a chiral model is assumed for hadrons and the same parameter values assumed in both plots. The comparison indicates that both the neutron and mercury EDM bounds are overestimated by the original {\tt CPsuperH} for the assumed parameter values. 

\end{document}